\documentclass[sigconf,nonacm,9pt]{acmart}
\usepackage{physics}

\usepackage{cancel}
\usepackage{booktabs,multirow}
\usepackage[font=small,skip=4pt]{caption}
\usepackage{stfloats}
\usepackage[ruled,vlined,linesnumbered]{algorithm2e}
\usepackage{amsmath}

\usepackage{amssymb}

\renewcommand{\authorsaddresses}{}
\begin{document}
\title{Quantum Circuit Synthesis Using an Exact T Library}
\author{Hanyu Wang}
\affiliation[obeypunctuation=true]{\institution{University of California, Los Angeles}, \city{Los Angeles}, \country{USA}}
\author{Mingfei Yu}
\affiliation[obeypunctuation=true]{\institution{\'Ecole Polytechnique F\'ed\'erale de Lausanne}, \city{Lausanne}, \country{Switzerland}}
\author{Xinrui Wu}
\affiliation[obeypunctuation=true]{\institution{University of California, Los Angeles}, \city{Los Angeles}, \country{USA}}
\author{Jason Cong}
\affiliation[obeypunctuation=true]{\institution{University of California, Los Angeles}, \city{Los Angeles}, \country{USA}}

\begin{abstract}
In fault-tolerant quantum circuit synthesis, T gates supplied via magic states dominate space–time cost, while Clifford gates incur negligible overhead. Conventional flows minimize AND count in an \{XOR, AND, NOT\} basis as a proxy for T, which neglects phase cancellation and can be far from T-optimal. We instead formulate an exact T synthesis problem and canonicalize Boolean functions under Clifford equivalence. By precomputing T-optimal implementations up to seven variables and developing a customized mapper, we reduce the T count by up to 14.3\% on EPFL benchmarks and improve the T counts of several cryptographic modules by up to 40\%.
\end{abstract}

\thanks{This work is supported by the NSF Challenge Institute for Quantum Computation (CIQC, Grant No. 2016245), the National Quantum Virtual Laboratory (NQVL, Grant No. 2410716), NSF Grant No. CCF-2313083, and a gift from Google LLC. J.C. has a financial interest in Google.
}

\maketitle

\section{Introduction}
Fault-tolerant quantum computing (FTQC) is becoming practical as systems reach scales where quantum error correction (QEC) sustains long-lived logical qubits. 
Growth in physical qubits, coherence, and control fidelity is pushing code distances to the regime needed for nontrivial algorithms, moving FTQC from aspiration to engineering~\cite{google2023suppressing, google2025quantum, sales2025experimental}.
In these architectures, T gates are the bottleneck resource and hinder us from running practical applications fault-tolerantly. 
This is because Clifford gates can be executed transversally (e.g., color codes), while non-Clifford gates such as T require magic-state distillation and injection with a 3D code~\cite{arute2019quantum, gidney2025factor, zhou2025resource}. 
The magic state factory's throughput and layout dominate the cost, despite the latest progress on state cultivation~\cite{litinski2019game, gidney2024magic, tan2024sat, bluvstein2024logical}.
Therefore, minimizing T count is critical.
\begin{figure}[t]
    \centering
    \includegraphics[width=\linewidth]{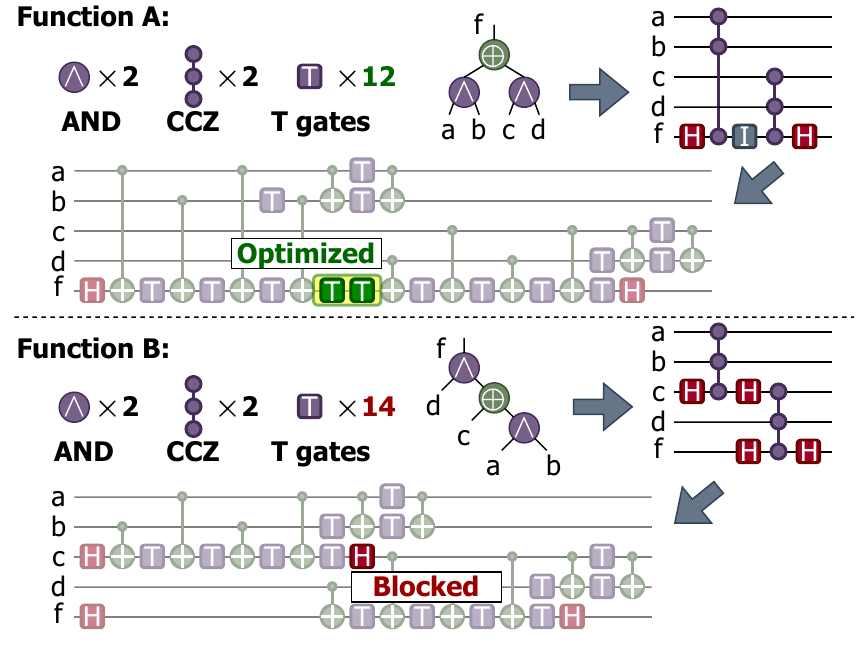}
    \Description{Two example circuits with identical AND count but different T cost.}
    \vspace{-0.7em}
    \caption{Example of the same AND count results in different T costs. Although both functions A and B contain two ANDs, the two Ts in A can be merged, while a Hadamard blocks those in B.}\label{fig:correlation}
    \vspace{-1em}
\end{figure}

When implementing modules, existing methods map Boolean functions onto the \{XOR, AND, NOT\} basis~\cite{halevcek2017xors} and decrease the number of AND gates, as they require Ts. 
Riener et al.~\cite{riener2019exact} introduces a SAT-based method that guarantees the minimum number of minterms in a two-level XOR-AND expression.
Meuli et al.~\cite{meuli2019evaluating} evaluates the gap between terms and AND count, and develops a weighted objective.
Testa et al.~\cite{testa2019reducing} proposes affine-equivalence and minimizes the AND count in multi-level XOR-AND graphs (XAG).
Meuli et al.~\cite{meuli2020ros} extends the library to $k$-LUT mapping. 

Although the AND count reflects the T cost, classical and quantum flows measure ``nonlinearity'' differently. 
AND count is only a proxy; it often fails to predict non-Clifford cancellations and phase simplification and can be far from T-optimal. 
For instance, functions A and B in Figure~\ref{fig:correlation} have the exact same AND count, but Ts in function A can be optimized by canceling two adjacent Ts highlighted, while function B cannot due to the Hadamard (see caption). 
Indeed, once XAG mapping is fixed, subsequent quantum circuit optimizations cannot easily recover missed structural opportunities due to the inserted Hadamard gates and allocated ancilla qubits.
As a result, T optimizations, e.g., AlphaTensor~\cite{ruiz2025quantum} and Fast\textsc{TODD}~\cite{vandaele2024lower}, cannot further improve these circuits. 
Therefore, assuming each AND uniformly transforms to a fixed number of Ts is not only inaccurate but 
also leads to an irreversible loss in optimality. 

In this work, we refine the formulation on XAG to directly use the exact T count to drive optimizations and mapping. By canonicalizing Boolean functions under Clifford equivalence, we precompute a library of T-optimal implementations for functions with up to seven variables and customize a mapping algorithm for circuit optimization.
Our approach combines the strengths of existing works: we keep the graph topology of the XOR-AND graph~(XAG) to preserve structural information for technology mapping, and we introduce a quantum-aware cost model that reads the exact T count via phase polynomial representation. Compared to XAG mapping driven by exact AND count, our method produces higher-quality Clifford+T circuits; next to standalone T optimizations applied after mapping, ours offers better scalability and circuit quality by exploiting the underlying logic network.

On the \texttt{EPFL} arithmetic benchmarks, our mapper achieves up to $14.3\%$ T count reduction compared to prior works on XAG. 
After integrating into the circuit synthesis flow, we lower the best known T count for running important cryptography applications by up to 40\%, demonstrating that preserving XAG structure while reasoning with exact T cost yields tangible improvements.

\section{Classical and Quantum Nonlinearity}\label{sec:background}
As mentioned, all existing works use the AND count, i.e., multiplicative complexity~(MC), as the proxy for T count optimization. 
This is because they measure nonlinearity in classical and quantum logic, respectively. 
In order to illustrate their differences in Section~\ref{sec:tensor}, we introduce their connections as background in this section.

\subsection{Multiplicative Complexity of Classical Functions}
The nonlinearity of a Boolean function $f:\{0,1\}^{n}\to\{0,1\}^m$ is captured by its MC: the minimum number of AND gates in any XOR-AND graph (XAG), in which XOR is linear (treated as free) and AND measures nonlinearity.

The MC of a Boolean function can be canonically upper-bounded by its \emph{algebraic normal form} (ANF). ANF counts nonlinear monomials $S$ with the corresponding coefficients $a_{S}$:
\[
f(x_1,\dots,x_n)=\bigoplus_{S\subseteq[n]} a_S\prod_{i\in S}x_i,
\]
whose na\"ive realization builds each monomial of degree $|S|\ge2$ with a tree using $|S|-1$ ANDs and XORs the results, giving
\[
\text{MC}_{\text{ANF}}(f)=\sum_{\substack{S\subseteq[n],\,a_S=1}}\bigl(|S|-1\bigr).
\]

For example, Figure~\ref{fig:bilinear}(a) contains an XAG with inputs $a,b,c,d$ and outputs $f$ and $g$. Nodes in the XAG are ANDs (purple) and XORs (green), and edges are wires or inverters (dashed). We derive the ANF of $f=\neg (a\wedge (\neg b))$ using the property $\neg x=1\oplus x$:
\begin{equation}\label{eqn:anf-example}
    f = 1\oplus a (1\oplus b) = 1\oplus a\oplus ab,
\end{equation}
from where we read the AND cost $0\!+\!0\!+\!1\!=\!1$. 

This bound is often loose: sharing, negation, and affine rewriting can reduce ANDs. For example, $h= ab\oplus abc$ requires $2$ instead of $1\!+\!2\!=\!3$ ANDs because the term $ab$ can be shared. 

Therefore, prior work canonicalizes functions up to affine equivalence (e.g., via Walsh-Hadamard spectra)~\cite{meuli2019role} and synthesizes \emph{exact} MC libraries available up to about 5-6 inputs~\cite{testa2019reducing}.
However, these works still fail to consider sharing between outputs. For instance, although $f$ and $g$ in Figure~\ref{fig:bilinear} require one and three ANDs, respectively, $ab$ can be shared. This dual-output function requires at most $3$, instead of $1\!+\!3\!=\!4$ ANDs.

\begin{figure}[b]
    \centering
    \vspace{-0.5em}
    \includegraphics[width=\linewidth]{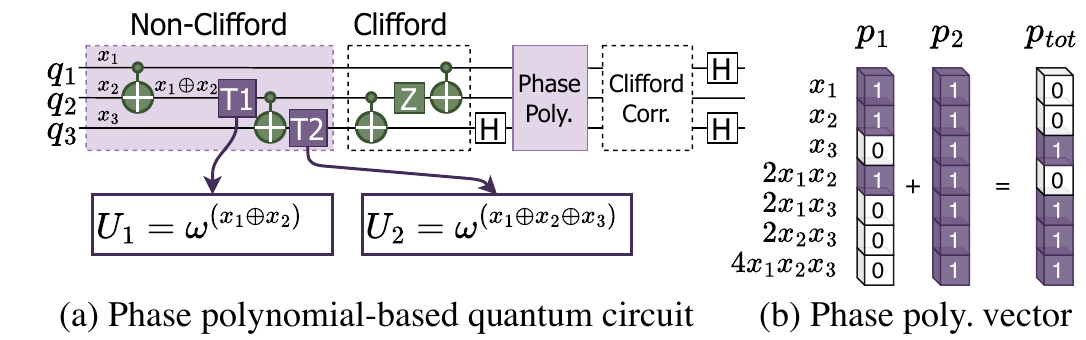}
    \Description{Circuit with two T gates and corresponding phase polynomial and vector representations.}
    \vspace{-0.6em}
    \caption{A circuit with two Ts ($T_1$, $T_2$) and two representations of the phase shift on $\ket{x_1x_2x_3}$: (a) phase polynomial and its (b) vector form.}
    \label{fig:phase}
    \vspace{-1em}
\end{figure}

\subsection{Non-Clifford Cost of Quantum Functions}
The nonlinearity of a quantum circuit arises from its non-Clifford gates, e.g., T gates. As in Figure~\ref{fig:phase}, a circuit can be viewed as alternating \emph{phase layers} and Hadamard gates~\cite{vandaele2022phase}. Each phase layer, 
\[
\ket{x}\;\longmapsto\; \omega^{f(x)}\ket{A x},
\]
applies a linear basis change $A$ and a $\omega=e^{i\frac{\pi}{4}}$ that encodes the function $f(x)$ on the exponent of the phase terms, which requires non-Clifford gates to implement.

Non-linearity of function is computed as the multiplication on the \emph{phase polynomial} $f$.
Each T gate contributes one term to $f(x)$, with CNOTs defining the parities. For instance, the two T gates in Figure~\ref{fig:phase}~(a) induce parities $x_1\!\oplus\!x_2$ and $x_1\!\oplus\!x_2\!\oplus\!x_3$, respectively. 

These terms will introduce nonlinear phase shifts $f(x)$ on the state $\ket{x_1x_2x_3}$, according to the phase-state duality~\cite{amy2021phase}.
\begin{align*}
    f(x)&=(x_1\oplus x_2)+(x_1\oplus x_2\oplus x_3) \\
    &=(x_1+x_2-2x_1x_2)+(x_1+x_2+x_3-2x_1x_2-2x_1x_3 \\
    &\quad-2x_2x_3+4x_1x_2x_3) \\
    &=x_3-2x_1x_3-2x_2x_3+4x_1x_2x_3 + 2(x_1+x_2-2x_1x_2),
\end{align*}
where the first four terms are T gates but the last three are two times the T, i.e., S gates. and are free. 
We use the vector form shown in Figure~\ref{fig:phase}~(b) to represent the contribution of each T gate, where each entry corresponds to a monomial and is set to 1 if that monomial appears an odd number of times in the polynomial.

\section{Canonicalize Bilinear Functions under Clifford Equivalence}\label{sec:tensor}
\begin{figure}[b]
    \centering
    \includegraphics[width=\linewidth]{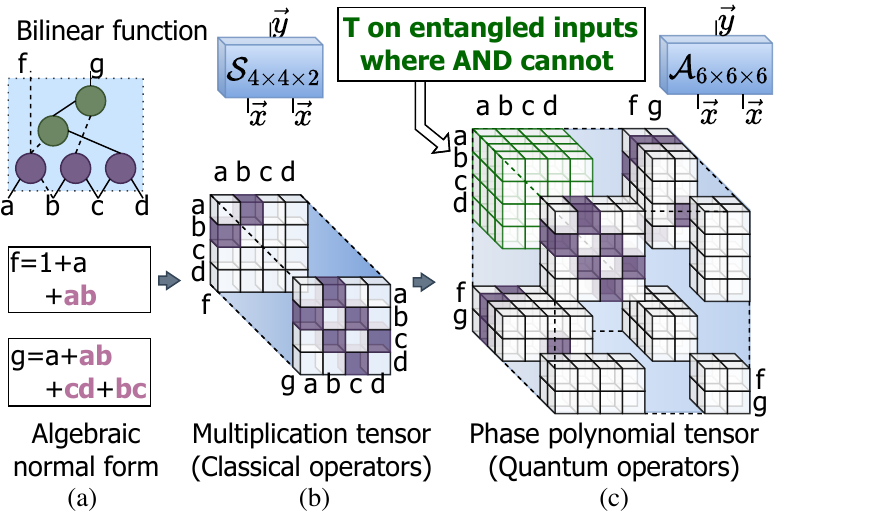}
    \Description{Tensor representation of a bilinear Boolean function, comparing classical multiplication tensor and phase polynomial tensor.}
    \vspace{-0.5em}
    \caption{Tensor representation for bilinear Boolean functions. We present an example with a function with $n=4$ inputs and $m=2$ outputs. Algebraic normal form (ANF) is derived according to Equation~(\ref{eqn:anf-example}). Compared to the classical multiplication tensor, the phase polynomial tensor incorporates the entries that entangle inputs and outputs (highlighted in green). T can operate in this highlighted corner, where a classical bilinear operator cannot.}
    \label{fig:bilinear}
    \vspace{-1em}
\end{figure}

While MC accurately quantifies nonlinearity in the classical context, it diverges from the true cost metric in the quantum setting, namely, the minimum T count required for a Clifford+T implementation. 

In this section, we introduce a canonical form of \emph{bilinear functions} under \emph{Clifford equivalence} to facilitate the exact T formulation. 

\textbf{Bilinear function.}
An $n$-input, $m$-output bilinear function $f : \{0,1\}^n \to \{0,1\}^m$ has its inputs partitioned into two disjoint subsets $U$ and $V$ with $U \cup V = [n]$. The $k$-th output can be written as
\[
y_k(x)\;=\;\bigoplus_{(i,j)\in U\times V} x_i x_j,
\]
which implies that $y_k$ admits an ANF of degree at most two. We will use the two-output function in Figure~\ref{fig:bilinear}(a) as a running example to illustrate how classical and quantum representations lead to different problem formulations, namely, exact MC versus exact T count, and how these formulations differ.

\textbf{Multiplication tensor}. 
Classically, we use \emph{multiplication tensor}, as displayed in Figure~\ref{fig:bilinear}(b), to canonicalize Bilinear functions under affine equivalence. 
Quadratic coefficients in a degree-2 ANF can be collected into an $n\times n \times m$ tensor $\mathcal{S}$. 
Each nonzero entry in the tensor represents an AND operation; more specifically, $\mathcal{S}_{i,j,k} = 1$ if the $x_i\wedge x_j$ contributes to the output $y_k$.
For example, given the expression in Figure~\ref{fig:bilinear}, the multiplication tensor has the nonzero triplets: $(a,b,f), (a,b,g), (c,d,g)$, and $(b,c,g)$. 
This classical form only couples input pairs with the output channel. 
\begin{figure}[t]
    \centering
    \includegraphics[width=\linewidth]{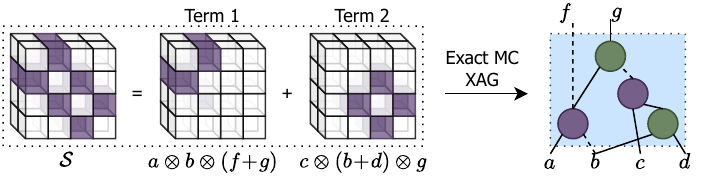}
    \Description{Rank-2 decomposition of a multiplication tensor and the corresponding XAG.}
    \vspace{-2em}
    \caption{Example of a multiplication tensor. The MC is given by the number of terms to decompose the tensor $\mathcal{S}$. Here $\text{MC}=2.$}
    \label{fig:decompose}
    \vspace{-0.5em}
\end{figure}

The exact MC can be defined as the rank of this tensor $\mathcal{S}$~\cite{fawzi2022discovering}, i.e., the minimum $r$ such that we can decompose $\mathcal{S}$ as:
\begin{equation}
    \mathcal{S}=\sum_{i=1}^{r}u_i\otimes v_i\otimes w_i,
\end{equation}
where $u_i,v_i,w_i$ are the linear combinations of inputs and outputs. Figure~\ref{fig:decompose} shows the decomposition by which we lower the MC of the bilinear function in Figure~\ref{fig:bilinear}(a) from three to two. The first term indicates the sharing of $ab$ between $f$ and $g$, while the second term rewrites $bc\!+\!cd$ into $c(b\!+\!d)$. Each term corresponds to an AND in the resulting XAG displayed on the right of Figure~\ref{fig:decompose}, i.e., $\text{MC}=2$.

\textbf{Phase polynomial tensor}.
To convert the classical function into a quantum operator, we embed it reversibly on $N=n+m$ qubits via the compute-in-place map $(x,y)\mapsto(x,\,y\oplus f(x))$. 
In this reversible setting, the non-Clifford content is captured by an order-3 tensor $\mathcal{A}\in\{0,1\}^{N\times N\times N}$ that records triple-parity couplings among all wires, including both inputs and outputs together. 
Similar to the multiplication tensor, we can derive from $y=f(x)$:
\begin{equation}\label{eqn:tensor}
\mathcal{A}_{i,j,k} = 1 \Leftrightarrow x_i\wedge x_j \text{ contributes to } y_k,
\end{equation}
for inputs $x_i,x_j$ and output $y_k$, and the minimum number of Ts required to implement $\mathcal{A}$ is, similarly, the rank of $\mathcal{A}$~\cite{heyfron2018efficient}.

After extending $\mathcal{S}_{n\times n\times m}$ to $\mathcal{A}_{N\times N\times N}$ by padding zeros, quantum operations can entangle inputs and outputs, creating third-order interactions that move information both within the input space and across the input–output boundary. T gates and related non-Clifford gadgets operate in this enlarged tensor space, whereas classical bilinear operators cannot access it. By exploiting the extended tensor, quantum circuits can find ``shortcuts'' in the decomposition, so the resulting T cost can be lower than the estimate from multiplicative complexity, as will be illustrated in Section~\ref{sec:synth}.

This viewpoint encourages using the tensor $\mathcal{A}$ on the combined $(n+m)$-body system as the canonical form for reasoning about quantum cost, i.e., the exact T cost. In other words, phase polynomial canonicalizes Boolean functions under Clifford equivalence.

\section{Exact T Library Synthesis}\label{sec:synth}
Unlike prior work that proxies T cost by AND or multiplicative complexity, the goal of this work is to \emph{directly} compute the exact T count for a bilinear Boolean function. 
We first uniquely derive its tensor representation $\mathcal{A}_f$ via Equation~(\ref{eqn:tensor}), which can be written in its vector form $p_f$. To construct our exact T library, we must find a Clifford+T circuit that implements the phase polynomial vector exactly. In this section, we describe how these mappings are built and how the exact T library is synthesized.
\begin{table}[b]
    \small
    \vspace{-1em}
    \caption{Exact T synthesis variable declaration}\label{tab:variables}
    \vspace{-0.6em}
    \begin{tabular}{@{}ll@{}}
        \toprule
        \textbf{Variable} & \textbf{Description} \\
        \hline
        \textbf{Input:} \\
        $f:\{0,1\}^{n} \mapsto \{0,1\}^{m}$ & An $n$ -input $m$-output bilinear function \\
        $N\in\mathbb{N}$ & Number of qubits, $N=n+m$ \\
        $K\in\mathbb{N}$ & Number of T candidates, $K=2^N-1$ \\
        $M\in\mathbb{N}$ & Number of monomials, $M=\sum_{r=1}^{3}\binom{N}{r}$. \\
        \hline
        \textbf{Internal:} \\
        $p_f\in\{0,1\}^{M}$ & The phase polynomial of $f$\\
        $\Phi\in\{0,1\}^{M\times K}$ & Mapping from T to the phase polynomial\\
        $p_k\in\{0,1\}^{M}$ & The phase polynomial of $k$~th T candidate\\
        $v_i\in\text{ker}(\Phi)\subset \{0,1\}^K$ & $i$~th basis vector of $\text{ker}(\Phi)$, $i\in[K\!-\!\operatorname{rank}(\Phi)]$ \\
        $u_j\in\{0,1\}^K$ & $j$~th kernel utilizing clean ancilla\\
        \hline
        \textbf{Output:} \\
        $s_k\in\{0,1\}$ & Whether T candidate $k$ is selected \\
        $s\in\{0,1\}^{K}$ & The vector form of all $s_k$, i.e., $s=\{s_1,\dots,s_K\}$ \\
        \bottomrule
    \end{tabular}
\end{table}

\subsection{Exact T Synthesis Problem Formulation}\label{sec:problem-formulation}
We formulate the exact T synthesis problem as a constrained integer programming problem. 
All utilized variable declaration is listed in Table~\ref{tab:variables}, and we will detail them throughout this section.

\textbf{Input and output variables}. The input of our synthesis problem is the bilinear function $f$ with $n$ inputs and $m$ outputs. We need to allocate $N=n+m$ qubits for $f$. 

A CNOT+T circuit with $N$ qubit has $K=2^N-1$ possible T candidates, each corresponding to the parity of a set of qubits excluding the empty set. 
We have to decide which T candidates to select to correctly implement the function $f$. 
The output is a vector $s$ that indicates whether candidate T is selected ($s_k\!=\!1$) in the circuit.  

\textbf{Functionality constraints}. To facilitate the linear programming problem formulation, we use the phase vector representation introduced in Figure~\ref{fig:phase} to describe circuit functionality.
According to Equation~\ref{eqn:tensor}, we can derive the phase polynomial tensor $\mathcal{A}_f$ by setting the triplets $(i,j,k)$ to $1$.
The target tensor $\mathcal{A}_f$ can be transformed to a phase polynomial vector $p_f$ by mapping triplet coordinates $(i,j,k)$ to the corresponding monomial. 

\begin{figure}[t]
    \centering
    \vspace{-1em}
    \includegraphics[width=\linewidth]{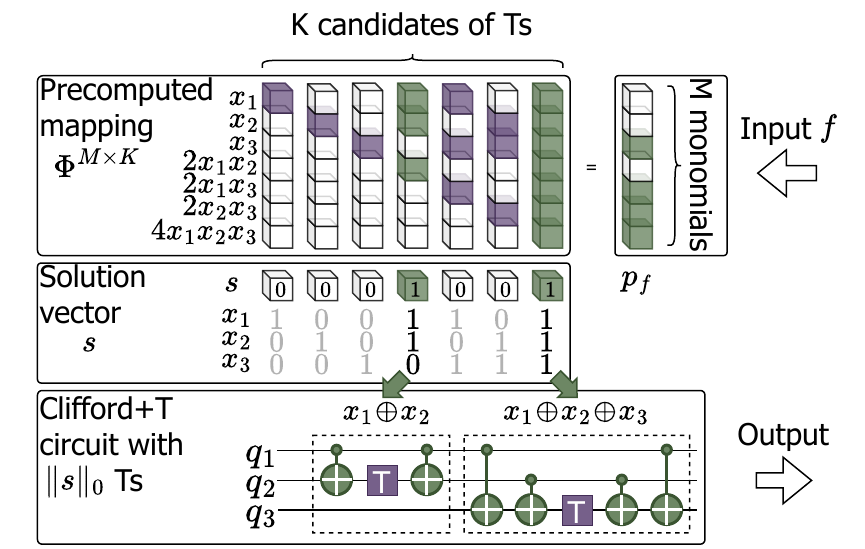}
    \Description{Exact T synthesis example showing the linear system between T candidates and target phase polynomial.}
    \vspace{-0.5em}
    \caption{Exact T synthesis example. The solver finds $s$ with minimum $\|s\|_0$ satisfying $\Phi \cdot s=p_f\; (\mathrm{mod}\;2)$, where $\Phi$ maps parities to monomials. Selected vectors are highlighted in green.}
    \label{fig:synth}
    \vspace{-0.5em}
\end{figure}

Instead of $N^3$, it suffices to use $M=\sum_{r=1}^{3}\binom{N}{r}$ dimensions in $p_f$ to represent $\mathcal{A}_f$ due to the symmetry. This is because each non-Clifford phase is uniquely determined by the support set $S\subseteq[N]$ of the parity on which a T gate acts; its contribution depends only on $S$, not on the ordered triplet $(i,j,k)$; and parities with $|S|\ge 3$ are Clifford-equivalent to combinations of smaller subsets~\cite{amy2019rmcode}, so only subsets of size at most $3$ yield independent coefficients.

Besides the target function, we also use vectors to represent the phase polynomial contributed by each T candidate. 
This is achieved by a precomputed linear map, $\Phi\in\{0,1\}^{M\times K}$, where column $k$ of $\Phi$ corresponds to the phase polynomial $p_k$ of the $k$~th T candidate.

When a T candidate $k$ is selected in a circuit, i.e., $s_k=1$, it contributes individually and linearly to the entire phase polynomial. 
Therefore, the functionality correctness requires the sum of phase polynomials $p_k$ contributed by Ts to be equal to the target polynomial $p_f$, i.e.,:
\begin{equation}\label{eqn:functionality-constraints}
    \Phi\cdot s = \sum_k s_k\cdot p_k = p_f \; (\mathrm{mod}\;2),
\end{equation}
where $s\in\{0,1\}^K$ is the solution vector that collects all the T candidates' selection decisions.

\textbf{Optimization objective}. Our goal is to minimize the number of T gates in the circuit, which is the cardinality of the selection vector $s$, $||s||_0 = \sum_k s_k$. Combining with Equation~(\ref{eqn:functionality-constraints}), we have:
\begin{equation}\label{eqn:problem-formulation}
\min ||s||_0
\quad \text{s.t.} \quad 
\Phi\cdot s = p_f \; (\mathrm{mod}\;2).
\end{equation}
This is a 0-1 optimization over parity constraints. Solving the constrained optimization problem in Equation~(\ref{eqn:problem-formulation}) is well-studied~\cite{muller1954application,amy2019rmcode}.

Figure~\ref{fig:synth} shows the case $N=3$, where all $K=2^{N}-1=7$ nonzero candidates appear as columns of $\Phi$. The target is $p_f=x_3+2x_1x_3+2x_2x_3+4x_1x_2x_3$ in the example in Figure~\ref{fig:phase}. The solution $s$ has parity terms $k\!=\!x_1\!\oplus\!x_2$ and $k\!=\!x_1\!\oplus\! x_2\!\oplus\!x_3$ selected and others not. $s$ is feasible since the XOR of corresponding vectors equals $p_f$. This solution is optimal because no single column matches $p_f$, so $\|{s}\|_0=1$ is impossible.

After solving the optimal solution $s$, we construct the Clifford+T circuit by transforming each $k$ into a phase gadget whenever $s_k=1$. The gadget comprises CNOT gates to prepare the parity term and a single T gate to implement the phase shift. Therefore, the returned circuit has exactly $\|{s}\|_0$ number of Ts. 

\subsection{Don't Care Conditions from Clean Ancilla}
The formulation in Equation~(\ref{eqn:problem-formulation}) treats all T-candidates uniformly and enforces only the phase constraints $\Phi \cdot s = p_f \pmod 2$. 
However, among all $N$ qubits, only the first $n$ qubits are logical inputs that carry information when entering the circuit; the remaining $m$ qubits are introduced solely to store the outputs.
These $m$ qubits can be considered as \emph{clean ancilla} because they can be initialized in the ground state $\ket{0}$. 
For each output wire $k \in \{n+1, \dots, n+m\}$, we write $f_k(x)$ for its Boolean value on input $x = (x_1,\dots,x_n)$, and the final state of that qubit is
\[
    \ket{y_k} = \ket{f_k(x)}.
\]
Because the ancilla are now set to $\ket{0}^{\otimes m}$, we only necessitate the circuit to be correct on the subspace where the ancilla take these prescribed values; phases on other basis states have no observable effect. These are, therefore, don't-care conditions.
This freedom enables cost-free Clifford rewrites to further reduce T count.

As a simple example, consider an output wire with
\[
    f_3(x) = x_1 x_2,
\]
so the circuit maps a clean ancilla $q_3$ from $\ket{0}$ to $\ket{x_1 x_2}$. 
Applying an $S$ gate (a Clifford gate equal to two T gates in series) to $q_3$ in the state $\ket{x_1 x_2}$ introduces the phase
\begin{equation}
    2y_3 = 2x_1 x_2 = x_1 + x_2 - (x_1 \oplus x_2),
\end{equation}
with $x_i, y_i \in \{0,1\}$ and arithmetic over the integers. 

This means the same phase can be generated utilizing three T gates on the three parities on the right-hand side. Thus, the Clifford $S$ gate is functionally equivalent to these three T gates on all valid inputs. Because Clifford operations are free in our cost model, replacing these Ts by a single $S$ gate reduces the circuit's T count while preserving correctness.

This rewrite corresponds to adding a vector for each $j\in [m]$:
\begin{equation}\label{eqn:dont-care}
    u_j \in \{0,1\}^K,\quad \Phi \cdot u_j = 0 \pmod 2,
\end{equation}
that toggles exactly those three T candidates whose combined phase equals the single $S$ on the clean ancilla. 

Adding $u_j$ to any feasible selection vector $s$ produces another functionally correct solution $s\oplus u_j$, potentially with a smaller T count. Clean ancilla therefore introduce a \emph{don't care} condition, enlarging the set of phase-equivalent realizations by introducing additional nullspace directions.

\subsection{Solving a Linear System with an Extended Null Space}\label{sec:null-space}
For the case in Figure~\ref{fig:synth}, where $N=3$ and $M=K=7$, solving Equation~(\ref{eqn:problem-formulation}) exactly is trivial. 
Indeed, if the $\Phi$ is full rank and reversible, then the solution vector $s=\Phi^{-1}p_f$ is unique and, by definition, optimal. However, since the number of T candidates $K$ grows asymptotically faster than the possible phase polynomials $M$, the $\Phi$'s rank is lower than the dimension of $s$:
\begin{equation}
    \operatorname{rank}(\Phi)\leq M=\Theta(N^3),\qquad \dim(s)=K=2^N-1=\Theta(2^N).
\end{equation}

Meanwhile, clean ancillas contribute additional kernel vectors $u_j$ according to Equation~(\ref{eqn:dont-care}), which further enlarge $\text{ker}(\Phi)$. 
This expansion is beneficial because a larger null space expands the likelihood that the affine family $s\oplus\text{ker}(\Phi)$ contains lower-cardinality solutions with significantly fewer T gates. However, as the number of feasible solutions grows exponentially with the nullity of $\Phi$, exhaustive search, although necessary for an exact solution, becomes extremely time-consuming. 

\begin{algorithm}[t]
\small
\caption{\textsc{LibrarySynthesis}$(N, \mathcal{F})$}
\label{alg:library}
\KwIn{$N$ (qubit count) $\mathcal{F}$ (bilinear functions)}
\KwOut{$\mathcal{L}$: map $f \mapsto s$}
$\Phi \leftarrow \text{PrecomputeMap}(N)$; \, $V \leftarrow \text{NullSpaceBasis}(\Phi)$\;
\ForEach{$f \in \mathcal{F}$}{
  $p_f \leftarrow \text{PhasePoly}(f)$; \, $W_f \leftarrow V \cup \text{DCVecs}(f)$\;
  $s \leftarrow \text{GaussianSolve}(\Phi,p_f)$\;
  \ForEach{$u \in \mathrm{span}_{\mathbb{F}_2}(W_f)$}{
    $\hat{s} \leftarrow s\oplus u$\;
    \If{$\|\hat{s}\|_0 < \|s\|_0$}{$s \leftarrow \hat{s}$\;}
  }
  $\mathcal{L}[f] \leftarrow s$\;
}
\Return $\mathcal{L}$
\end{algorithm}

To harness the complexity and reduce the minimum cardinality of the solution vector, we develop an efficient library synthesis algorithm listed in Algorithm~\ref{alg:library} that precomputes $\Phi$ and a null-space basis $V$ and stores them in memory to speed up search. Moreover, we write a CUDA kernel to parallelize the search over candidate null-space vectors by assigning them to different threads. Our solver returns exact solutions for $N \le 7$ and suboptimal ones for $N > 7$. In practice, this is sufficient given the scalability of the cut-enumeration algorithm (see Section~\ref{sec:mapping}).

\section{Customized Library Mapping}\label{sec:mapping}
\begin{figure}[b]
    \centering
    \includegraphics[width=\linewidth]{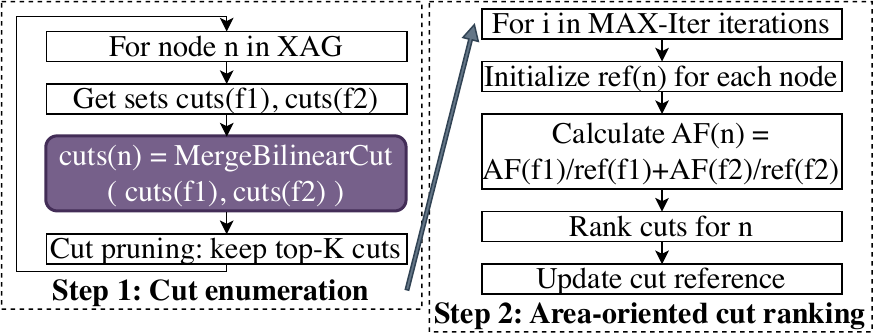}
    \Description{Customized technology mapping flow that enumerates bilinear cuts and queries the exact T library.}
    \vspace{-1.5em}
    \caption{Customized mapping flow with bilinear cut enumeration.}
    \label{fig:cut-mapping}
    \vspace{-1em}
\end{figure}

Our Exact T library supplies a quantum-aware cost model and exposes accurate T costs. The exact library avoids irreversible suboptimal choices for qubit allocation and Hadamard gate insertion during mapping. However, to fully benefit from the library, the mapping algorithm must be customized for it. Since the tensor formulation naturally targets bilinear functions, by prioritizing cuts that reveal bilinear structure, the mapper would better exploit the phase polynomial implementations in the Exact T library and achieve a lower overall T count.

The overall workflow of our customized mapping algorithm is depicted in Figure~\ref{fig:cut-mapping}. Our procedure follows the cut-based enumeration, pruning, and ranking flow in the VLSI design flow~\cite{chen2004daomap, cong1999cut}, but is customized to prioritize \emph{bilinear cuts}, i.e., cuts whose function is bilinear. 

\begin{figure}[t]
    \centering
    \vspace{-0.5em}
    \includegraphics[width=0.9\linewidth]{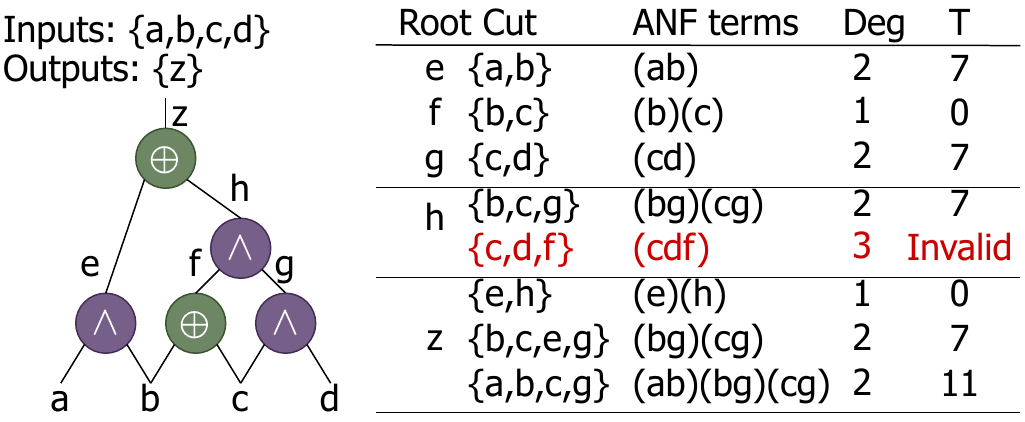}
    \Description{Bilinear cut enumeration on an XAG, pruning cuts whose ANF degree exceeds two.}
    \vspace{-0.5em}
    \caption{Bilinear cut enumeration example. Our flow keeps only cuts with ANF degree $\le 2$ so they admit a phase-polynomial representation. The highlighted cut (degree 3) is pruned.}
    \label{fig:mapping}
    \vspace{-0.5em}
\end{figure}

Figure~\ref{fig:mapping} demonstrates an example of our bilinear cut enumeration. The XAG before mapping has inputs $a, b, c, d$, output $z$, and internal nodes $e,f,g,h$. During mapping, we derive feasible cuts for each node as a root in topological order by merging enumerated cuts at the node's fanins. To keep the cut bilinear, we merge cuts differently depending on the type of the root:

\textbf{Node is AND} ($e,g$ and $h$). If \emph{both} fan-in ANFs are degree~1 (purely linear), we form their cross-products to obtain degree-2 terms (bilinear). E.g., cut $\{b,c,g\}$ rooted at $h$ has degree-2 terms $(bg)(cg)$ as the product of $(b)(c)$ and $(g)$. If \emph{either} fan-in already contains degree-2 terms on this cut, we \emph{prune} the cut, since the new AND would introduce degree~3 monomials not supported by our library (e.g., the cut $\{c,d,f\}$ rooted at $h$).

\textbf{Node is XOR} ($f$ and $z$). If any side contributes degree-2 terms, we drop the resulting degree-1 debris and keep only degree-2 terms. E.g., cut $\{e,f,g\}$ rooted at $z$ merges degree-2 terms $(fg)$ from $h$ and $(e)$, between which $(e)$ will be dropped. This ensures every retained cut remains degree~$\le 2$ and is bilinear.

Once the ANF terms are collected, we encode each quadratic monomial $x_i x_j$ as a triplet $(i,j,r)$, where $r$ denotes the cut root. The set of all such triplets defines the exact $T$-synthesis instance from Section~\ref{sec:synth} for that cut, from which we obtain its exact $T$ cost. The mapper then proceeds with standard area-oriented dynamic programming over cuts (Figure~\ref{fig:cut-mapping}), utilizing these T-accurate scores to select implementations.

\begin{table}[b]
\small
\centering
\caption{Benchmark families and parameters.}
\vspace{-0.7em}
\label{tab:benchmarks}
\begin{tabular}{l c l}
\toprule
\textbf{Family} & \textbf{\#} & \textbf{Parameters} \\
\midrule
EPFL arithmetic~\cite{amaru2015epfl} & 6 & MC-optimized XAGs~\cite{liu2022don} \\
Random QLUT oracles~\cite{venere2024quantum} & 36 & $n_{\rm in}\!=\!4$--$7$, $n_{\rm out}\!\in\!\{1,2,4\}$ \\
ModExp for Shor/RSA~\cite{gidney2019windowed} & 6 & $(m,n)$ up to $(5,221)$ \\
PrepareTHC for chemistry~\cite{babbush2018encoding} & 6 & $N_{\rm orb}\!=\!8$, varying precision \\
GF$(2^n)$ Mult for AES~\cite{luo2022quantum} & 7 & $n=2$--$8$ \\
\bottomrule
\end{tabular}
\vspace{-1em}
\end{table}

\section{Experimental Results}
This section evaluates our main contribution: using an Exact T library during mapping from an XAG to a Clifford+T quantum circuit. First, we compare against the existing XAG mapping flow to show that both the library alone and the customized mapping reduce the T count. Next, we integrate our approach into a complete quantum circuit synthesis flow and measure the end-to-end impact on application benchmarks. Finally, we analyze runtime, focusing on the overhead introduced during mapping.

\subsection{Experimental Setup}
We implement our algorithm in \texttt{C++}~\footnote{Available at: \url{https://github.com/Nozidoali/exact-t-map.git}}.
All experiments are run on an Apple M4 Max workstation with 36GB RAM; tool versions and configuration details appear in the documentation.
\textbf{Benchmarks.}
All used benchmarks are listed in Table~\ref{tab:benchmarks}. Our benchmark suites cover not only different functions but also different families, including arithmetic logic from EPFL benchmarks, random QLUT oracles, modular exponentiation for Shor's factorization algorithm, preparing blocks employing tensor hypercontrast for chemistry simulation, and $\mathrm{GF}(2^n)$ multipliers for AES S-Box. Each family spans multiple parameter settings and, where applicable, several randomized instances to test robustness across structures and applications.

\textbf{Baselines.}
We select the strongest baselines for comparison across the benchmark families. 
For EPFL arithmetic benchmarks, we compare the XAGs with the best-known MC results~\cite{liu2022don, yu2025back} and the mapping driven by exact MC~\cite{testa2019reducing}.
We use domain baselines for the end-to-end module synthesis comparison: AlphaTensor~\cite{ruiz2025quantum} and Fast\textsc{TODD}~\cite{vandaele2024lower} for $\mathrm{GF}(2^n)$ arithmetic, and Qualtran's QROM~\cite{babbush2018encoding} and SelectSwap~\cite{low2024trading} implementation for QROM-style oracles~\cite{harrigan2024expressing}. These baselines represent the best-known T counts for each benchmark, so any improvement over them is valuable and constitutes a meaningful advance for the field.

\textbf{Metrics.}
Since many baselines report only T but not Clifford or qubit counts, we compare space-time volumes according to Beverland's model~\cite{beverland2022assessing, harrigan2024expressing} only when they are available. Nevertheless, T gates dominate the cost in contemporary fault-tolerant architectures, and on our classical-circuit benchmarks, Clifford and T counts tend to follow the same trend.

\begin{figure}[t]
    \centering
    \vspace{-0.8em}
    \includegraphics[width=\linewidth]{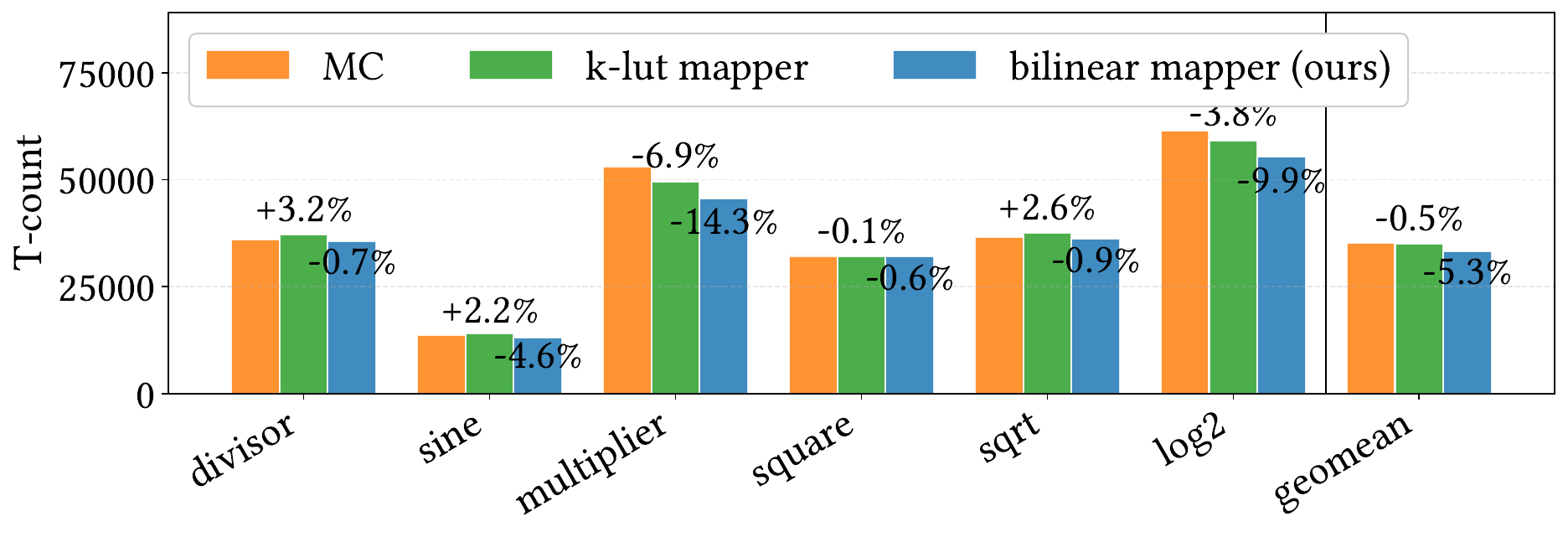}
    \Description{Bar chart comparing T counts of different mapping flows on EPFL arithmetic benchmarks.}
    \vspace{-0.5em}
    \caption{T-count (\(\downarrow\)) comparison on arithmetic benchmarks~\cite{amaru2015epfl}. Percentages above bars are reduction vs.\ MC.}\label{fig:exp-epfl}
\end{figure}

\begin{figure}[b]
    \centering
    \vspace{-1em}
    \includegraphics[width=\linewidth]{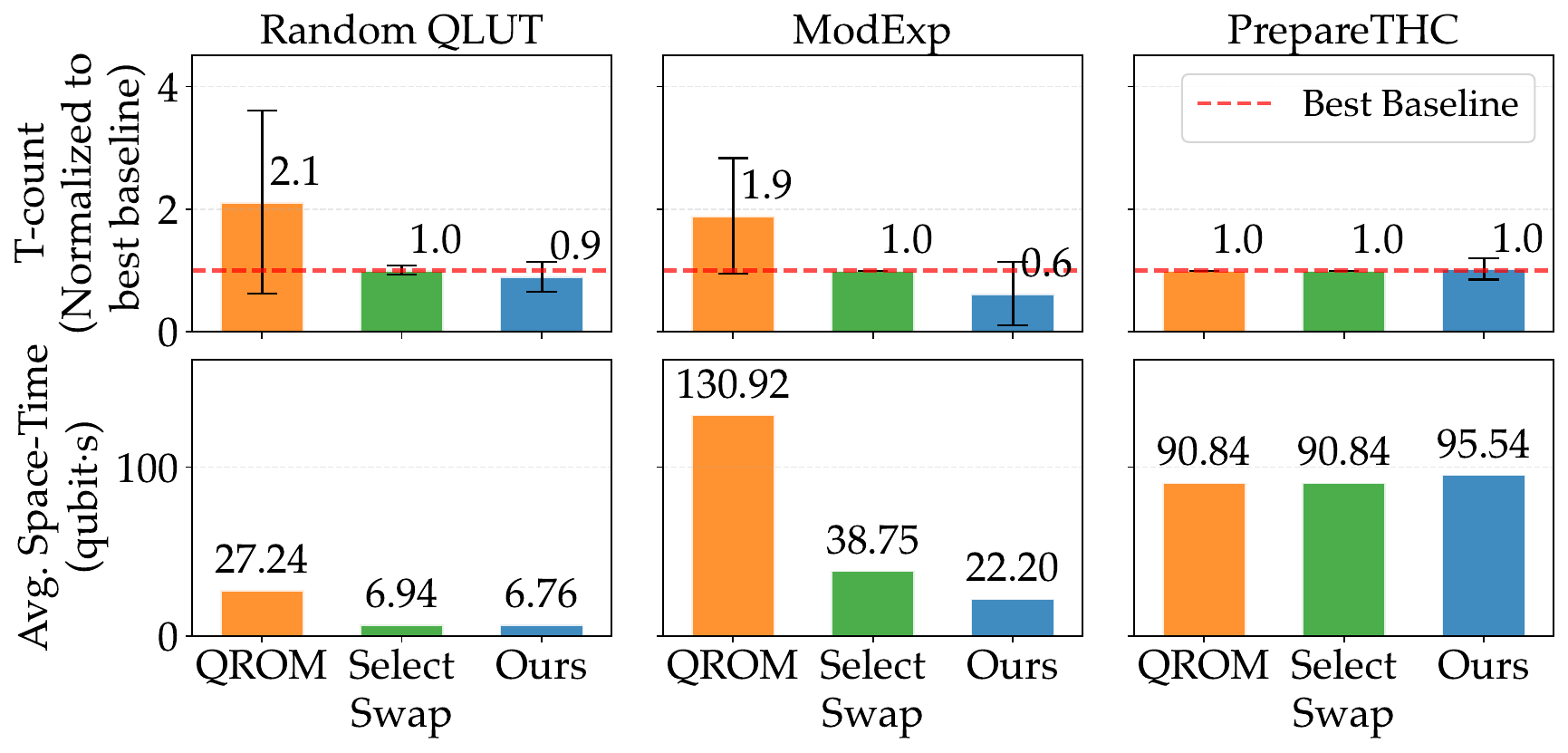}
    \Description{Normalized T count and space-time volume reduction on QLUT-based applications versus QROM and SelectSwap baselines.}
    \vspace{-0.0em}
    \caption{T count (\(\downarrow\)) and space-time volume (\(\downarrow\)) reduction on QLUT-based real-world applications compared with QROM~\cite{babbush2018encoding} and SelectSwap~\cite{low2024trading} implementations in Qualtran~\cite{harrigan2024expressing}. We normalize results to the best baseline on each benchmark before aggregation.}\label{fig:exp-qlut}
\end{figure}

\subsection{Mapping Comparison}
Figure~\ref{fig:exp-epfl} shows that replacing exact MC estimates with our exact T costs improves mapping quality. Even with the same $k$-Cut mapper~\cite{calvino2023technology}, our library yields about a 0.5\% geomean reduction. On benchmarks with richer bilinear structure, such as \texttt{multiplier}, the T count reduction increases to 14.3\%. However, because cut ranking and pruning are unaware of the library’s strengths, some advantageous structures may be discarded by the heuristics, leading to quality degradation on a few benchmarks. Overall, the library supplies more accurate T costs and steers cut selection toward better decisions, which in turn translates into lower T counts after mapping.

We next enable our customized mapping algorithm. This keeps bilinear structure explicit during cut enumeration and ranking rather than obscuring it behind local heuristics. As seen in Figure~\ref{fig:exp-epfl}, when equipped with this customized mapper, our Exact T library achieves T count results that dominate existing methods on all benchmarks. On average, we reduce T count by 5.3\% (geomean), an order of magnitude larger than the gains obtained with the conventional $k$-Cut mapper.

\begin{figure}[t]
    \centering
    \includegraphics[width=0.8\columnwidth]{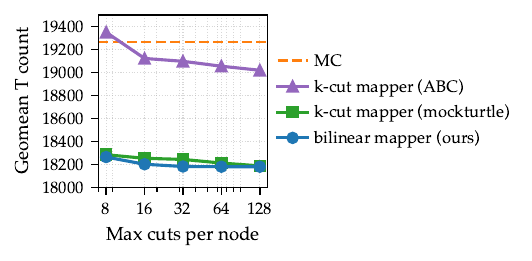}
    \Description{Geometric mean T count of three mappers as the cut-pruning budget varies.}
    \caption{Geomean T-count (\(\downarrow\)) of three mappers as the cut-pruning budget varies on six EPFL benchmarks. The dashed line is the multiplicative complexity-based analytical reference.}\label{fig:mcpn-ablation}
\end{figure}

Figure~\ref{fig:mcpn-ablation} sweeps the cut-pruning budget on three mappers that share our Exact T library, including \texttt{amap} from ABC~\cite{brayton2010abc}, \texttt{emap} in Mockturtle~\cite{calvino2023technology}, and our mapper. Our customized mapper saturates at 8 cuts per node, where expensive per-cut solver is still affordable, whereas the mockturtle $k$-Cut mapper still improves up to about 64 cuts per node before approaching the same plateau. The ABC $k$-Cut mapper never matches the quality of the other two because XOR is not a native node in its AIG, so exploiting the free XORs that our library relies on is much harder.

\subsection{End-to-End Circuit Synthesis Comparison}
We next place the mapper inside a full quantum-circuit synthesis flow and measure its end-to-end impact. Since our method takes an XAG as input and produces a Clifford+T circuit, integrating it into a complete flow requires two additional subroutines:

\textbf{Synthesis} extracts an XAG network from the specification, including lookup tables and arithmetic expressions. We apply a two-level decomposition using ANFs to \textsc{prepare} all the minterms~\cite{yu2025anf} and positive Davio expansion to \textsc{select} from minterms~\cite{ko2002efficient}.

\textbf{Resynthesis} iteratively applies peephole optimization, extracting local windows from the XAG and rewriting them into alternative structures with lower cost. In each iteration, we apply graph balancing~\cite{haner2020lowering}, rewriting~\cite{testa2019reducing}, and resubstitution~\cite{wang2023anysyn} until convergence (no further MC reduction in an iteration).

\begin{figure}[b]
    \centering
    \vspace{-0.5em}
    \includegraphics[width=\linewidth]{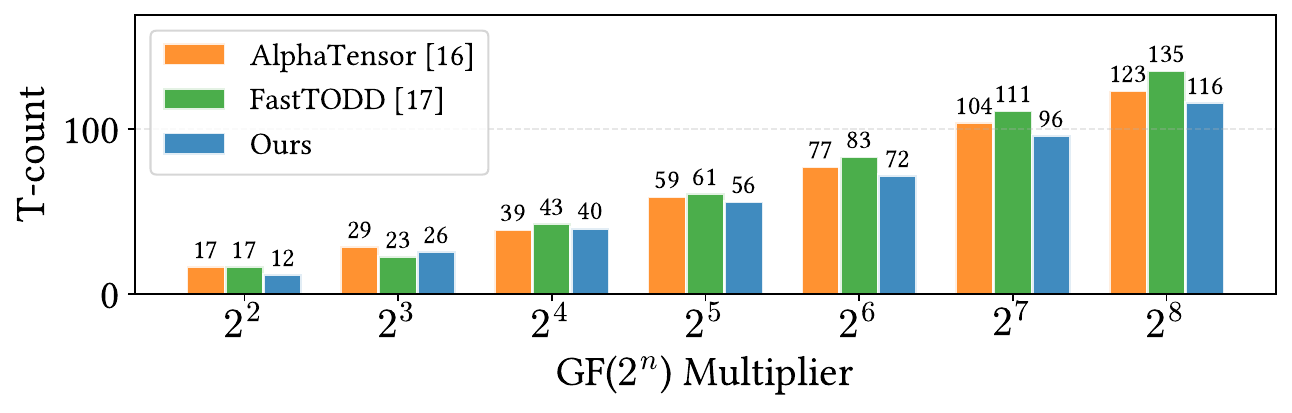}
    \Description{Bar chart of T counts on Galois-field multipliers compared with prior implementations.}
    \vspace{-0.5em}
    \caption{T-count (\(\downarrow\)) comparison on GF(\(2^n\)) multipliers~\cite{luo2022quantum}.}\label{fig:gf-mult}
\end{figure}

Figure~\ref{fig:gf-mult} and Figure~\ref{fig:exp-qlut} summarize the comparison on Galois-field multipliers and QLUT-based real-world applications, respectively. Because the internal representation preserves XAG structure throughout the flow, we obtain consistent T count and space-time volume reductions on QLUT oracles and modular exponentiation. For QLUTs with extremely high entropy, such as \textsc{prepareTHC} blocks, our results are comparable, with further gains largely limited by the quality of the initial XAG synthesis algorithm.

For $\mathrm{GF}(2^n)$ multipliers, our flow achieves lower T counts than prior implementations on most bitwidths. This is enabled by don’t-care conditions exposed at the logic-network level, which allow us to change the unitary in ways that prior works, e.g., AlphaTensor~\cite{ruiz2025quantum} and FastTODD~\cite{vandaele2024lower}’s formulations, cannot. Overall, keeping an XAG at the core of the flow, pairing it with an exact-T library, and preserving structure during mapping yields tangible end-to-end T count savings beyond what mapping-only changes can deliver.

\subsection{Runtime Breakdown and Mapping Overhead}
\begin{figure}[t]
    \centering
    \vspace{-0.5em}
    \includegraphics[width=\linewidth]{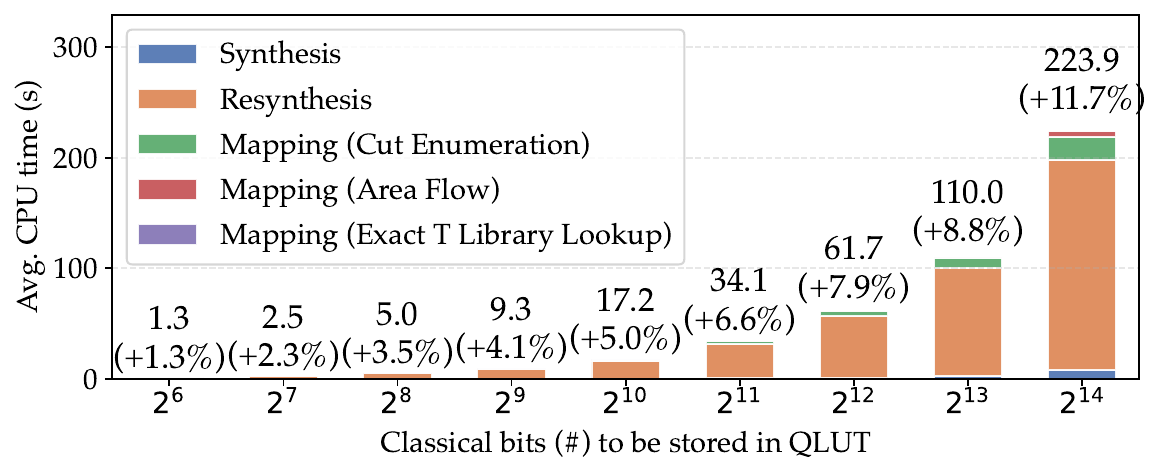}
    \Description{Stacked bar chart breaking CPU time into synthesis, resynthesis, and three mapping stages on QLUT benchmarks.}
    \vspace{-0.5em}
    \caption{CPU time breakdown on QLUT benchmarks as QLUT size grows. Our mapper introduces an overhead of at most $11.7\%$.}\label{fig:cpu-time}
    \vspace{-0.5em}
\end{figure}

Figure~\ref{fig:cpu-time} partitions total runtime into five components: synthesis, resynthesis, and three mapping stages introduced by our method: \emph{cut enumeration} (customized for bilinear function), \emph{area flow}, and \emph{exact T library lookup}. Before our work, mapping time was negligible, so the three additional bars reflect the extra computation we introduced during mapping. This overhead is measurable but remains a small fraction (at most $11.7\%$) of the total runtime once synthesis and resynthesis are included. More importantly, it buys a clear reduction in circuit cost by exposing exact T costs and preserving structure where the search matters. In practice, this synthesis overhead is a one-time cost for quantum module designs that are reused across experiments and applications, so the runtime is amortized while the circuit improvements persist.

\section{Conclusion}
In this paper, we present an exact T library for Clifford+T circuit synthesis. By introducing a canonical form for Boolean functions under Clifford-equivalence, our formulation targets T cost directly rather than relying on multiplicative complexity as a proxy. The library supplies accurate T costs during mapping. Combined with a customized mapper that emphasizes bilinear functions during cut pruning and ranking, our approach outperforms existing flows and attains T counts substantially below the best-known results. 

\balance
\bibliographystyle{abbrv}
\bibliography{references}
\end{document}